# POWER SPECTRUM ANALYSIS OF MOUNT-WILSON SOLAR DIAMETER MEASUREMENTS: EVIDENCE FOR SOLAR INTERNAL R-MODE OSCILLATIONS

P.A. STURROCK[1] AND L. BERTELLO[2]

[1] Center for Space Science and Astrophysics, Varian 302, Stanford University, Stanford, CA 94305; sturrock@stanford.edu

[2] National Solar Observatory, 950 North Cherry Avenue, Tucson, AZ 85719

## ABSTRACT

This article presents a power-spectrum analysis of 39,024 measurements of the solar diameter made at the Mount Wilson Observatory from 1968.670 to 1997.965. This power spectrum contains a number of very strong peaks. We find that eight of these peaks agree closely with the frequencies of r-mode oscillations for a region of the Sun where the sidereal rotation frequency is 12.08 year$^{-1}$. We estimate that there is less than one chance in $10^6$ of finding this pattern by chance.

*Subject heading:* Sun: oscillations; Sun: photosphere; methods: data analysis; statistical

1. INTRODUCTION

This article is based on a compilation of solar diameter measurements made at the Mount Wilson Observatory. The file comprises 39,024 measurements made over the time interval 1968.670 to 1997.965, approximating 3.65 measurements per day. In this article, dates are quoted in a format introduced for the analysis of solar neutrinos: in order to obtain a smoothly running set of dates that are close to calendar dates, we count days from 1970 January 1 as day 1, and then convert to years by dividing by 365.2564 and adding 1970.

In this article, we are concerned with r-mode oscillations (Papaloizou and Pringle, 1978; Provost, Berthomieu, and Rocca, 1981; Saio, 1982). These are retrograde waves in a rotating fluid with frequencies determined by the sidereal rotation frequency, $\nu_R$, and the values of $l$ and $m$, two of the three spherical harmonic indices. The values of $l$ and $m$ are restricted by

$$l = 2,3,...,\ m = 1,2,...,l. \qquad (1)$$

The frequencies, as they would be measured by an observer co-rotating with the Sun, are given by

$$\nu(l,m,S) = \frac{2m\nu_R}{l(l+1)}. \qquad (2)$$

The corresponding frequencies, as they would be measured by an observer on Earth and when expressed in cycles per year, are given by

$$\nu(l,m,E) = m(\nu_R - 1) - \frac{2m\nu_R}{l(l+1)}. \qquad (3)$$

In a recent analysis of Super-Kamiokande solar neutrino data (Sturrock, 2008a), we found evidence for a pattern of peaks in the power spectrum that may be interpreted in terms of r-mode oscillations. We found a set of five peaks corresponding to frequencies given by Equation (3) with m = 1 and l = 2, 3, 4, 5, and 6, for the rotation frequency 13.97 year$^{-1}$. We found that it is quite unlikely that this combination of peaks would arise by chance. The purpose of this article is to carry out a similar analysis of the Mount Wilson solar diameter data. We carry out the same analysis in Section 2, a modified analysis in Section 3, and a significance test in Section 4. We discuss the results in Section 5.

## 2. E-TYPE R-MODE FREQUENCIES

The diameter measurements have mean value 1919.58 arc seconds, and standard deviation 0.84 arc seconds. The measurements, normalized to mean value unity, are shown in Figure 1. We have formed the power spectrum by a likelihood procedure (Sturrock et al., 2005a) that is equivalent to the Lomb-Scargle procedure (Lomb, 1976; Scargle, 1982). The resulting power spectrum, restricted to the frequency range 0—20 year$^{-1}$, is shown in Figure 2. The biggest peak (not shown, with power 1500) is—not surprisingly—one at 0.08 year$^{-1}$, attributable to the solar cycle.

We now analyze the diameter data in the same way that we analyzed the Super-Kamiokande data, by combining the power spectra formed from the values of the power at the E-type r-mode frequencies (Equation (3)). As in our Super-Kamiokande analysis, we consider $m = 1$, $l = 2,3,4,5,6$. We combine the powers by means of the "combined power statistic" (CPS; Sturrock et al., 2005b), defined as follows:

If $U$ is the sum of $n$ powers,

$$U = S_1 + S_2 + ... + S_n , \qquad (4)$$

then the CPS is defined by

$$G(U) = U - \ln\left(1 + U + \frac{1}{2}U^2 + ... + \frac{1}{(n-1)!}U^{n-1}\right). \qquad (5)$$

This statistic has the following property: If each power is drawn from an exponential distribution (as is appropriate if the power is formed from a time series dominated by normally distributed random noise; Scargle, 1982), then $G$ also conforms to an exponential distribution.

This statistic is shown, as a function of the sidereal rotation frequency $v_R$, in Figure 3. Unlike the pattern found in our analysis of Super-Kamiokande data, there is no single dominant frequency. Hence there is no evidence from the analysis of E-type r-mode frequencies for an associated group of r-modes.

## 3. S-TYPE R-MODE FREQUENCIES

We now repeat the analysis of Section 2, except that we now consider the S-type frequencies, given by Equation (2), retaining the same set of indices: $m = 1$, $l = 2,3,4,5,6$. The combined power statistic formed from these five powers is shown, as a function of the sidereal rotation frequency, in Figure 4. We see that, in this case, there is a clearly dominant peak, the frequency of which is 12.08 year$^{-1}$. Since the CPS has the same exponential distribution as a power, the peak is much more sharply defined than is apparent in this figure, due to the very large range in the statistic. For instance, we find that the 4-sigma range, for which the CPS drops by a factor of $10^8$, is 12.04 to 12.12.

Figure 5 shows the power spectrum, with arrows indicating the peaks corresponding to the S-type r-mode frequencies for m = 1, $v_R = 12.08$ and $l = 2,3,4,5,6,7,8,10$. The green arrows denote the frequencies of the five peaks (for $l = 2,3,4,5,6$) that contributed to the combined power statistic shown in Figure 4. The red arrows denote the (predicted) frequencies corresponding to the same value of $v_R$ for $l = 7,8,10$. There is no peak corresponding to $l = 9$, but the corresponding frequency (0.27 year$^{-1}$) is very close to that of the strong peak at 0.22 year$^{-1}$, so it may be subsumed in the wing of the latter peak. We see from Figure 5 and from Table 1 that there is excellent correspondence between the peaks corresponding to the S-type r-mode frequencies for $v_R = 12.08$ and actual peaks (with the exception of the missing or hidden peak at 0.27 year$^{-1}$).

## 4. SIGNIFICANCE ESTIMATE

We now look for a procedure for obtaining a robust estimate of the significance of the result found in previous sections. A familiar test for obtaining a significance estimate of a peak in a power spectrum is the "shuffle test" in which one retains actual values of the measurements and of the times of the measurements, but randomly reassigns the two sets of numbers (Bahcall and Press, 1991). This gives an estimate of the probability of obtaining the actual power from a featureless time series, such as one dominated by normally distributed random noise. However, we see from Figure 2 that this test would be inappropriate for our purposes, since the time series contains many periodicities in addition to those of special interest.

Our goal is to estimate the chance of finding a prescribed pattern in the power spectrum—specifically finding the set of peaks (shown in Figure 5) related to r-mode frequencies. In order to avoid confusion with the major peak at 0.08 year$^{-1}$, associated with the solar cycle, we again restrict attention to the peaks determined by $m = 1$ and $l = 2,3,4,5,6$.

There is a clear trend in the distribution of power as a function of frequency: for the range of frequency shown in Figure 5, the height of peaks tends to vary inversely with frequency. In order to give approximately equal weight to the r-mode peaks, we therefore form the "frequency-weighed power" Z, defined by

$$Z = v \times S , \qquad (6)$$

where $S$ is the power. We then form the sum of this quantity for a given set of $l$-value frequencies:

$$H = Z_2 + ... + Z_n . \qquad (7)$$

We may now obtain a significance estimate by means of a variation of the "shake test," which we recently introduced as an alternative to the shuffle test (Sturrock *et al.*, 2010a). In the original version of the shake test, we retain a set of measurements but introduce small random displacements of the times associated with those measurements. In the present version, we retain the computed value of $Z$ for each mode but shift the frequencies by random amounts, expressible as

$$\nu_n \rightarrow \nu_n + D\nu \qquad (8)$$

where $\nu_2,...,\nu_6$ are the frequencies determined by Equation (2) for $\nu_R = 12.08 \text{ year}^{-1}$, $m = 1$, and $l = 2,3,...,6$, and $D\nu$ denotes a randomly selected frequency shift for each mode. This process is has been repeated 100,000 times. For current purposes, it was convenient to select $D\nu$ values from a uniform random distribution over the range $-0.2$ to $0.2 \text{ year}^{-1}$, but the results are insensitive to the precise range adopted. The resulting distribution of 100,000 estimates of the quantity H is shown in histogram form in Figure 6, and as a logarithmic display in Figure 7. We see that, according to this analysis, there is about one chance in $5 \, 10^6$ of obtaining the actual value of $H$ (1772.9) or more by chance.

We have repeated this calculation, considering all nine frequencies (for $l = 2,3,...,10$). The maximum value of $H$ is then found to be 2525.2, at exactly the same frequency, 12.08 year$^{-1}$. A repeat of the Monte Carlo calculation shows that there is about one chance in $5 \, 10^6$ of obtaining this value of $H$ (2525.2) or more by chance. The reason that this result is close to that obtained from the five frequencies (for $l = 2,3,...,6$) is that the width of the additional peaks is comparable to the separation between the peaks, so that the shake test is not sensitive to the additional peaks. This suggests that the shake test is, if anything, a conservative significance estimator.

## 5. DISCUSSION

From the excellent correspondence of the frequencies, from the powers of the eight peaks, and from the analyses of Sections 3 and 4, it is clear that the Mount Wilson diameter measurements yield very strong evidence for a cluster of r-mode oscillations in the solar interior. These peaks correspond to S-type frequencies rather than E-type frequencies. It is worth noting, in this context, that the famous Rieger periodicity, with a period of approximately 154 days (Rieger et al., 1984), occurs at the frequency 2.37 year$^{-1}$. This may be interpreted as an *S*-type r-mode frequency for the values $l = 3, m = 1, v_R = 14.23 \text{ year}^{-1}$ (Sturrock et al., 1999). According to estimates of the internal rotation frequency derived from helioseismology (Schou et al., 1998), this is the frequency at normalized radius 0.71, which places it in the tachocline. Hence it is not unreasonable to suppose that the group of r-mode oscillations we have identified in diameter measurements may originate somewhere in the solar interior where the sidereal rotation frequency is 12.08 year$^{-1}$. This would place it below the radiative zone, in or near a slowly rotating solar core.

Our recent analysis of nuclear decay-rate data acquired at the Physikalisch-Technische Bundesanstalt in Germany (Siegert et al., 1998), which includes a re-analysis of decay-rate data acquired at the Brookhaven National Laboratory (Alburger et al., 1986), shows that both datasets exhibit a strong periodicity at about 11.2 year$^{-1}$, which may be attributed to rotation at the sidereal frequency 12.2 year$^{-1}$ (Sturrock et al., 2010b). We may also compare these results with the outcome of a recent combined analysis (Sturrock, 2008b) of Homestake (Cleveland et al., 1998; Davis, 1996; Davis et al. 1968) and GALLEX (Anselmann et al., 1993, 1995) solar neutrino data and ACRIM (Willson, 1979, 2001) total solar irradiance measurements. This analysis yields strong evidence for a modulation, common to all three datasets, at 11.85 year$^{-1}$, which would correspond to a sidereal rotation rate of 12.85 year$^{-1}$.

It seems likely that these periodicities have their origin in or near the solar core, and that this region is not in rigid rotation so that different manifestations of this rotation may yield different values. [There have in fact been theoretical conjectures that the Sun may have a slowly rotating core (Talon *et al.,* 2002; Charbonnel and Talon, 2005).] Hence further study of neutrino data, irradiance data, diameter data, and decay data, may offer insight into the structure and properties of the deep solar interior. It is also to be expected that the analysis of helioseismology data now being acquired by the Solar Dynamics Observatory will in due course yield information concerning the deep interior.

An interesting theoretical challenge will be to understand why r-mode oscillations are excited, and why they should influence diameter measurements. A possible answer to the first question is that the modes are for some reason unstable, so that the amplitudes grow from noise level to finite values. We may now note that the Rayleigh-Taylor instability in fluid dynamics (Paterson, 1983) and the two-stream instability in plasma physics (Sturrock, 1994) both involve the interplay of waves in components that have different velocities. We have already noted that the Rieger oscillation may be attributed to an r-mode oscillation (with $l = 3, m = 1$) that develops in the tachocline, in which there is a sharp gradient in angular velocity. Hence it is possible that the r-mode oscillations that influence the solar diameter originate in a region separating the core from the radiative zone, where there appears to be a gradient in angular velocity.

The answer to the second question (why the r-mode oscillations in or near the core influence the solar diameter) is also challenging. A possible cause is that the oscillations excite some kind of wave that travels from the vicinity of the core to the photosphere. In the ideal model (that is spherically symmetric), r-mode oscillations do not involve density fluctuations and therefore could not involve sound waves. However, if there is a magnetic field in or near the core, r-mode oscillations would lead to perturbations in that field, possibly leading to MHD waves that could travel to the photosphere. It is also possible the nuclear burning plays a role in response to, or in causing, departures from spherical symmetry.

In view of the fact that processes responsible for r-mode oscillations in or near the core may involve departures from spherical symmetry, nuclear burning, and magnetic field, it seems likely that detailed understanding of these processes will require computer modeling.

We thank Ephraim Fischbach and Roger Ulrich for helpful discussions concerning this project. This work of PAS was supported in part by the National Science Foundation through grant AST-0097128.

TABLE CAPTION

Table 1. Comparison of best-fit r-mode frequencies to actual peaks in the power spectrum.

| $l$ | Fitted Frequency (year$^{-1}$) | Predicted Frequency (year$^{-1}$) | Actual Frequency (year$^{-1}$) | Power |
|---|---|---|---|---|
| 2 | 4.03 | | 4.04 | 69.5 |
| 3 | 2.01 | | 2.01 | 245.6 |
| 4 | 1.21 | | 1.21 | 497.5 |
| 5 | 0.81 | | 0.80 | 253.0 |
| 6 | 0.58 | | 0.57 | 405.1 |
| 7 | | 0.43 | 0.42 | 315.5 |
| 8 | | 0.34 | 0.35 | 126.8 |
| 9 | | 0.27 | | |
| 10 | | 0.22 | 0.22 | 541.9 |

FIGURES

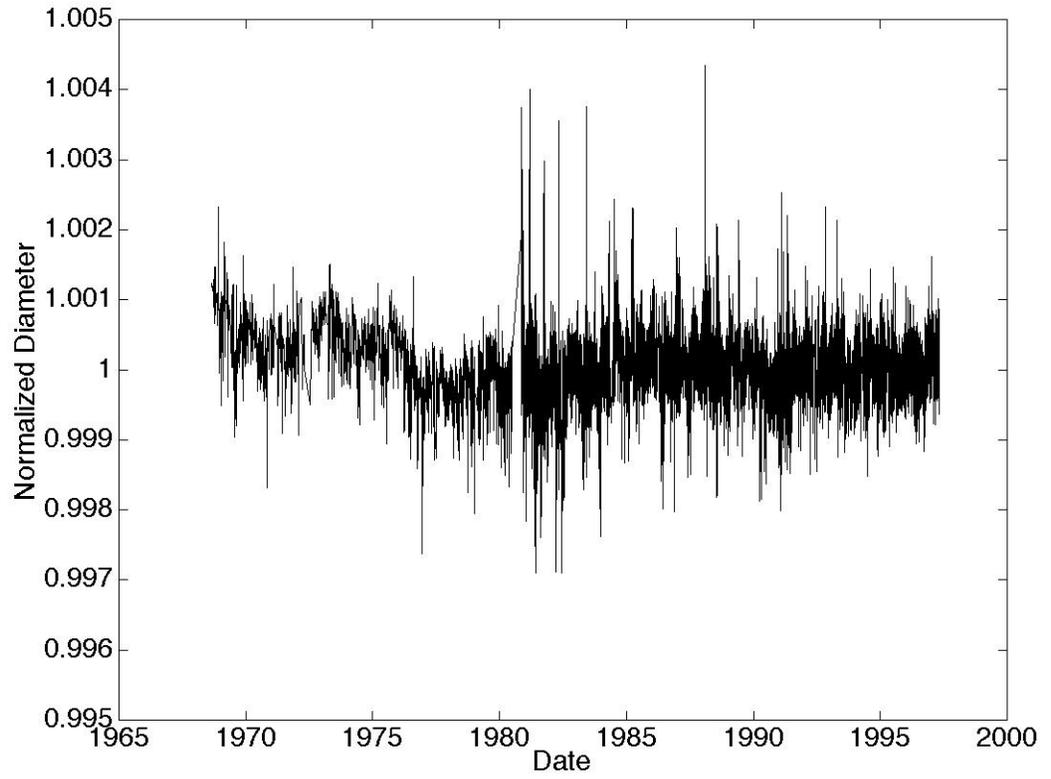

Figure 1. Normalized Mount Wilson solar diameter measurements.

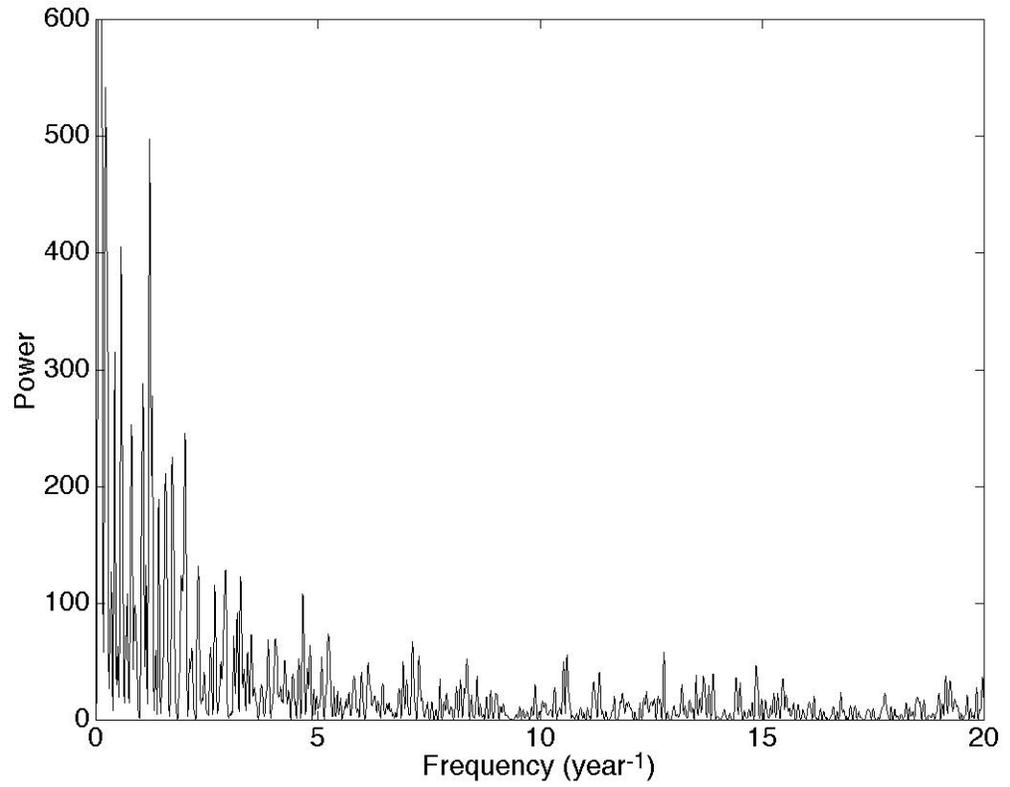

Figure 2. Power spectrum formed from the Normalized Mount Wilson solar diameter measurements.

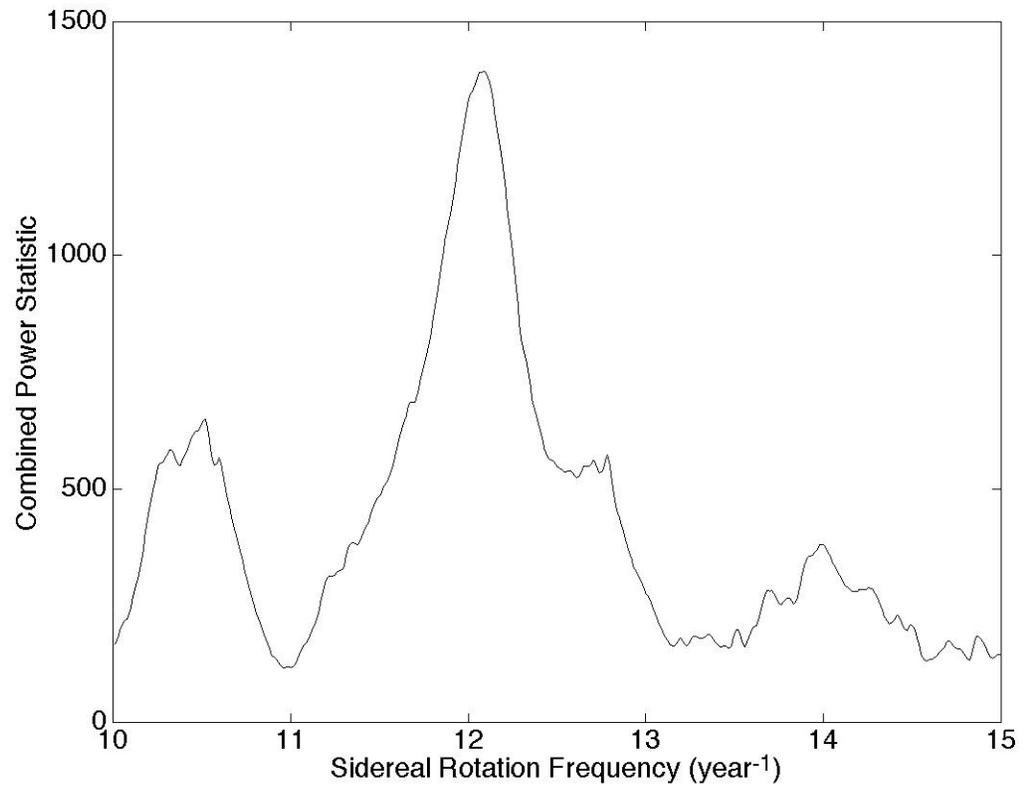

Figure 3. Combined power statistic formed from powers for the E-type frequencies corresponding to m = 1, l = 2,3,…,6.

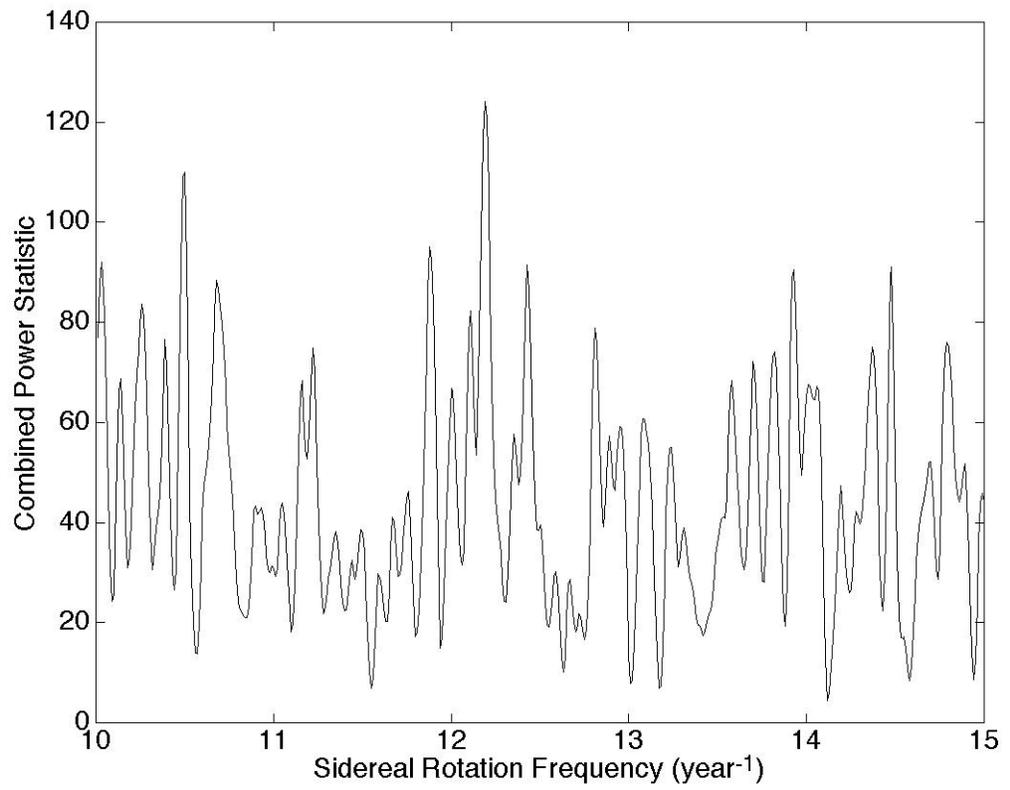

Figure 4. Combined power statistic formed from powers for the S-type frequencies corresponding to m = 1, l = 2,3,…,6.

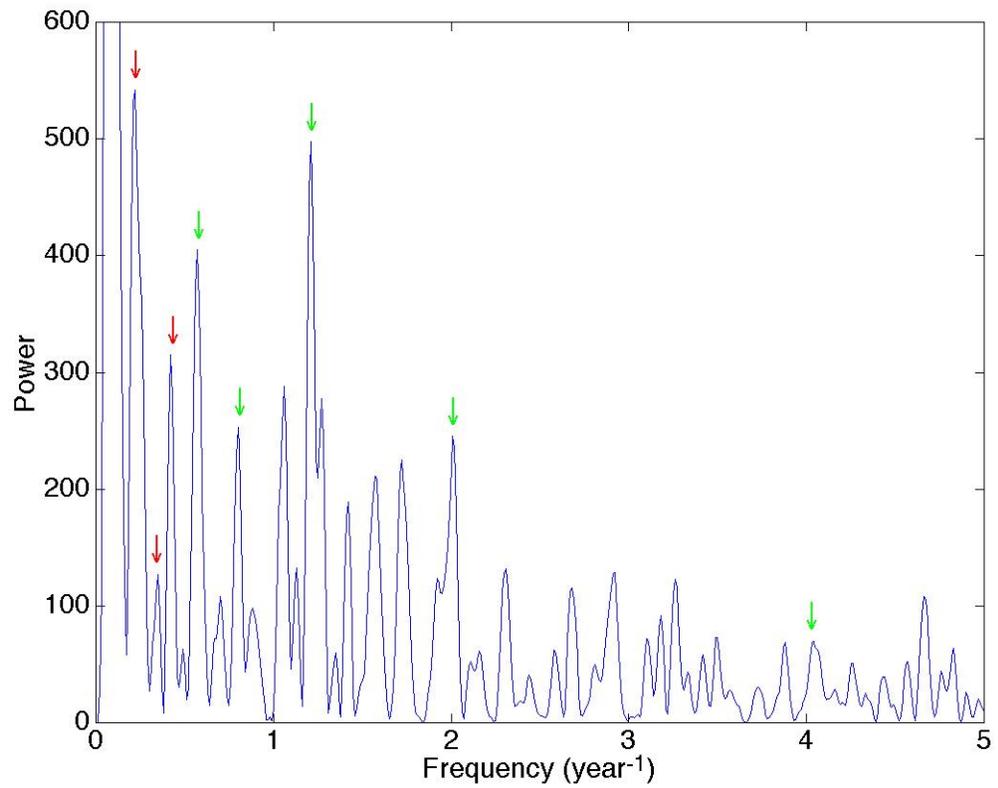

Figure 5. Power spectrum, showing in green the five peaks that contribute to the combined power statistic shown in Figure 4, and showing in red three additional peaks that did not contribute to the statistic shown in Figure 4.

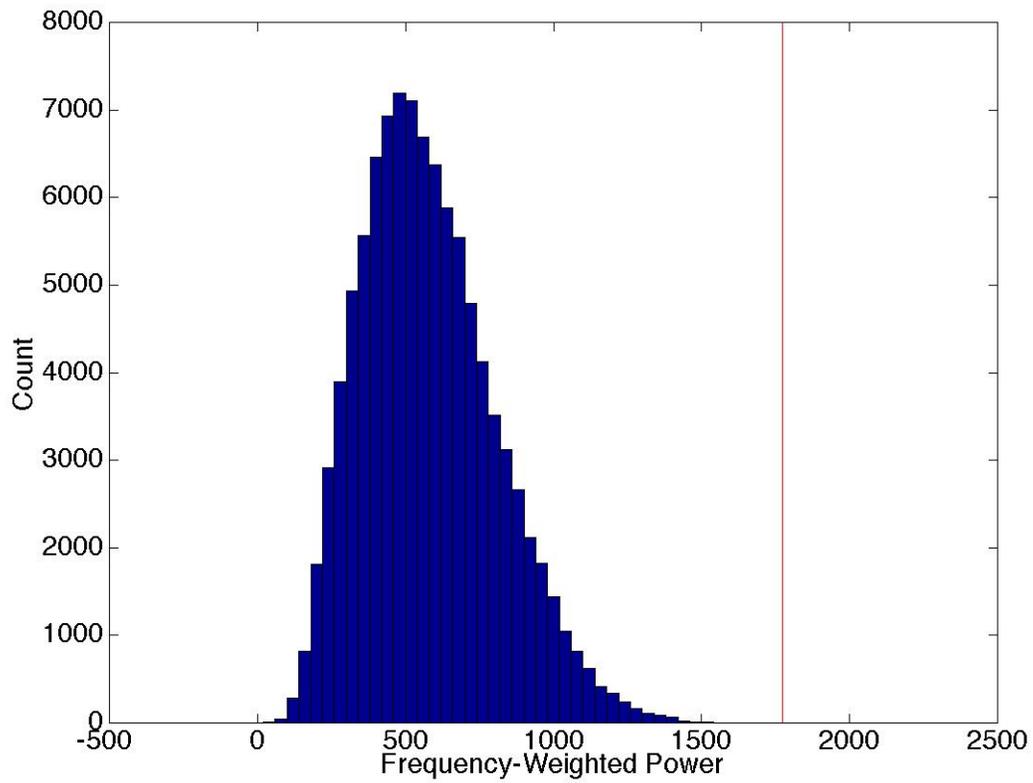

Figure 6. Histogram formed from 100,000 estimates of the sum of the frequency-weighted powers computed for $l = 2,...,6$, generated by the shake procedure. None is as large as that derived from the actual data (1772.9).

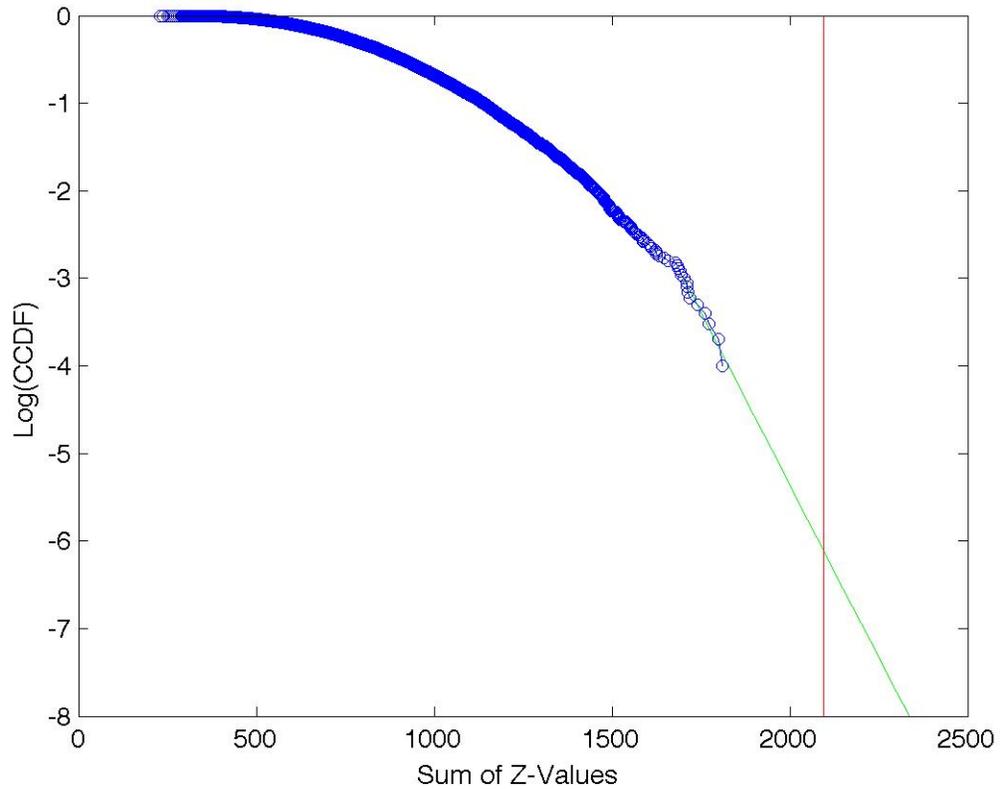

Figure 7. Logarithmic plot of the Complementary Cumulative Distribution Function (CCDF) of 100,000 estimates of the sum of the frequency-weighted powers computed for $l = 2,...,6$, generated by the shake procedure. A projection of the curve indicates that the probability of obtaining a peak value as large as or larger than the actual value (1772.9) is less than $10^{-6}$.